\documentclass[12pt]{article}
\usepackage{a4}
\usepackage{epsfig}

\newcommand{\DmsqS}{\Delta m^2_{\rm sol}}
\newcommand{\DmsqA}{\Delta m^2_{\rm atm}}
\newcommand{\DmsqAmS}{\Delta m^2_{32}}

\newcommand{\ket}[1]{|#1\rangle}

\newcommand{\Aa}{A}
\newcommand{\beq}{\begin{equation}}
\newcommand{\eeq}{\end{equation}}
\newcommand{\oddParam}{d}
\newcommand{\DMnsq}{\tilde{M}^2_{\nu}}

\def\ni{\noindent}
\def\nl{\hfill\break}
\begin{document}
\begin{flushright}
hep-ph/0403278 \\
RAL-TR-2004-012 \\
20 Apr 2004 \\
\end{flushright}
\begin{center}
{\Large The Simplest Neutrino Mass Matrix}
\end{center}
\vspace{1mm}
\begin{center}
{P. F. Harrison\\
Department of Physics, University of Warwick\\
Coventry CV4 7AL. UK\footnotemark[1]}
\end{center}
\begin{center}
{and}
\end{center}
\begin{center}
{W. G. Scott\\
CCLRC Rutherford Appleton Laboratory\\
Chilton, Didcot, Oxon OX11 0QX. UK\footnotemark[2]}
\end{center}
\vspace{1mm}
\begin{abstract}
\baselineskip 0.6cm
\noindent
We motivate the simplest ansatz for the neutrino mass matrix consistent
with the data from neutrino oscillation experiments, and admitting $CP$
violation. It has only two free parameters: an arbitrary mass-scale and a
small dimensionless ratio. This mass matrix exhibits two symmetries,
Democracy and Mutativity, which respectively ensure trimaximal mixing of the
$\ket{\nu_2}$ mass eigenstate, and mixing parameter values
$|\theta_{23}|=45^{\circ}$ and $|\delta|=90^{\circ}$, consistent with
bimaximal mixing of the $\ket{\nu_3}$ mass eigenstate. A third constraint
relates the smallness of $|U_{e3}|^2$ to that of the mass-squared
difference ratio, $\DmsqS/\DmsqA$, yielding the prediction
$\sin{\theta_{13}}=\sqrt{2\DmsqS/3\DmsqA}\simeq 0.13\pm 0.03$.
\end{abstract}

\begin{center}
({\it Submitted to Physics Letters B})
\end{center}

\footnotetext[1]{E-mail:p.f.harrison@warwick.ac.uk}
\footnotetext[2]{E-mail:w.g.scott@rl.ac.uk}

\newpage
\ni {\bf 1 Democracy and the Solar Data}
\vspace{2mm}
\nl The oscillation data from solar neutrino experiments \cite{solar} and from 
the KAMLAND experiment \cite{kamland} all point to a common conclusion,
namely that the magnitude, $|U_{e2}|$, of the relevant MNS 
\cite{mns} lepton mixing matrix element is given by $|U_{e2}|^2\simeq 1/3$.
Combined with information on atmospheric neutrino oscillations 
\cite{atmos}, from the Chooz and Palo Verde \cite{reactor} reactor
neutrino experiments and the constraints of unitarity, this strongly 
suggests \cite{characters} that the mixing matrix has one trimaximally 
mixed column corresponding to the $\nu_2$ mass eigenstate:
\begin{eqnarray}
\ket{\nu_2}=\frac{1}{\sqrt{3}}
\left( \matrix{  1 \cr
                 1 \cr
                 1 \cr } \right).
\label{nu2}
\end{eqnarray}
This, in turn, implies that the neutrino mass-squared matrix, 
$M^2_{\nu}:=M_{\nu}M^{\dag}_{\nu}$,
in the flavour basis\footnote{Working in the flavour basis, we choose the 
epsilon phase convention \cite{venice}, in which the freedom to rephase
the charged lepton flavour eigenstates is used to set the imaginary part 
of the neutrino mass-squared matrix proportional to the
totally anti-symmetric $3\times 3$ matrix, $\Aa$, defined by 
$\Aa_{\alpha \beta} := \epsilon_{\alpha \beta \gamma}$
($\gamma \neq \alpha , \beta$) $\alpha , \beta , \gamma = 1,2,3$.}
is an $S3$ group matrix \cite{characters} (ie.~a $3\times 3$ `magic square' \cite{venice}):
\begin{eqnarray}
M^2_{\nu}=
\left( \matrix{  x  &  z  &   y  \cr
                 z  &  y  &   x  \cr
                 y  &  x  &   z  \cr } \right) +
\left( \matrix{  a  &  b  &  b^* \cr
                b^* &  a  &   b  \cr
                 b  & b^* &   a  \cr } \right)
\label{s3group}
\end{eqnarray}
where $a$, $x$, $y$ and $z$ are real parameters and $b$ is complex
(we note that $M^2_{\nu}$ is proportional to the effective Hamiltonian for
neutrino oscillations, $H=M^2_{\nu}/2E$, where $E$ is the neutrino energy).

Such a mass matrix, Eq.~(\ref{s3group}), is the most general one which commutes with the Democracy operator \cite{venice}:
\begin{equation}
[M^2_{\nu},{\cal D}]=0
\label{commutator1}
\end{equation}
where the Democracy operator, ${\cal D}$, is defined as:
\begin{eqnarray}
{\cal D}=\frac{1}{3}
\left( \matrix{ 1  &  1 &  1  \cr
                1  &  1 &  1  \cr
                1  &  1 &  1  \cr } \right).
\label{democracy}
\end{eqnarray}
Hence, the neutrino mass eigenstates (displayed in the
Appendix, Eq.~(\ref{Umns})) are all eigenstates of ${\cal D}$,
with $\ket{\nu_2}$ having eigenvalue ${\cal D}=1$, while $\ket{\nu_1}$ and 
$\ket{\nu_3}$ each have eigenvalue ${\cal D}=0$. The Democracy
operator generates a group of unitary transformations,
$U_{\cal D}=e^{i\alpha{\cal D}}\,[=1+{\cal D}(e^{i\alpha}-1)]$ ($\alpha$ a real parameter), under which $M^2_{\nu}$ is invariant:
\begin{equation}
U_{\cal D}M^2_{\nu}\,U^{\dag}_{\cal D}=M^2_{\nu}.
\label{demInvariance}
\end{equation}
The usual consequences of unitary symmetry follow, eg.~the Democracy
quantum number is conserved in vacuum neutrino oscillations
(although, clearly, Democracy symmetry is broken by matter effects).

We note that our approach here is
distinct from that in eg.~Ref.~\cite{oldDemo} in which the mass matrix itself
was taken to be (approximately) proportional to ${\cal D}$ and was
then said to be (approximately) democratic. Here, Democracy refers to the
generator of the symmetry transformation, and matrices of the
(considerably more general) form, Eq.~(\ref{s3group}), are precisely
invariant under such transformations, Eq.~(\ref{demInvariance}).

The parameters used in Eq.~(\ref{s3group}) are not all observable, in the 
sense that, for a given set of observables, we may always add a given real 
constant to $a$ 
and $b$, providing that we subtract the same constant from $x$, $y$ and
$z$. This allows us to set ${\rm Re}(b)=0$, without loss of generality. 
Furthermore, only the neutrino mass-squared differences are observable
in neutrino oscillation experiments, so that the offset of the mass-squared 
eigenvalues is unobservable, and we may set the value of $a$ equal to 
zero in this context. The resulting reduced mass-squared matrix may then 
be written:
\begin{eqnarray}
\DMnsq:=MM^{\dag}-aI=
\left( \matrix{  x  &  z  &   y  \cr
                 z  &  y  &   x  \cr
                 y  &  x  &   z  \cr } \right) + i\frac{\oddParam}{\sqrt{3}}
\left( \matrix{  0  &  1  &  -1  \cr
                -1  &  0  &   1  \cr
                 1  & -1  &   0  \cr } \right)
\label{retro}
\end{eqnarray}
where $\oddParam=\sqrt{3}~{\rm Im}(b)$ is a measure of $CP$ violation.
We note that the two terms are respectively pure-real and pure-imaginary.
This simple four-parameter ansatz parameterises 
the six independent observables of neutrino oscillations: the solar and
atmospheric mass-squared differences, $\DmsqS$ and $\DmsqA$, the three mixing
angles \cite{PDG}, $\theta_{12}$, $\theta_{23}$, $\theta_{13}$ and
the $CP$-violating phase, $\delta$. Expressions for each of these are 
given in Eqs.~(\ref{s13sq})-(\ref{dmsqs}) of the Appendix.
Two degrees of freedom have been removed by the constraint of
Democracy invariance, which fixes the magnitudes of the
elements of the $\ket{\nu_2}$ mass eigenstate (only two of
which are independent),
Eq.~(\ref{nu2}).

\vspace{5mm}
\ni {\bf 2 Mutativity and the Atmospheric Data}
\vspace{2mm}
\nl 
The data on atmospheric neutrino oscillations \cite{atmos}, supported by
the K2K experiment \cite{k2k}, and from the Chooz and Palo Verde
\cite{reactor} experiments furthermore indicate that 
the $\nu_3$ mass eigenstate is approximately bi-maximally mixed between the 
$\mu$ and $\tau$ flavours. In the three generation context, exact bi-maximal
mixing of the $\nu_3$ eigenstate is understood to mean
$|U_{\mu 3}|=|U_{\tau 3}|=1/\sqrt{2}$,
or equivalently, $|U_{\mu 3}|=|U_{\tau 3}|$ {\bf and} $|U_{e3}|=0$.
Indeed, taking account of terrestrial matter effects \cite{messier},
which tend to decouple high-energy electron-neutrino mixing
at the atmospheric scale \cite{hps4}, it may be said that the data on
atmospheric neutrinos essentially require 
$|U_{\mu 3}|\simeq |U_{\tau 3}|$, while it is the reactor experiments, 
Chooz and Palo Verde, which require $|U_{e3}|^2 << 1$ (as will be 
addressed in the next section). Taken together, the combined data imply 
that the $\nu_3$ eigenstate is approximately bi-maximally mixed.

It is by now well-known \cite{mutausymm} that the first of the two
bi-maximal requirements, $|U_{\mu 3}|=|U_{\tau 3}|$, can be ensured by a form of $\mu-\tau$ permutation symmetry \cite{mutativity}. In the context of the ${\cal D}$-invariant ansatz, Eq.~(\ref{retro}), this may be enforced by simply setting $y=z$. It is shown in the Appendix that in the standard parameterisation,
this corresponds to fixing $|\sin{\theta_{23}}|=1/\sqrt{2}$,
and $|\sin{\delta}|=1$, exactly what is needed for
$|U_{\mu 3}|=|U_{\tau 3}|$. We note that for any fixed value of 
$\theta_{13}$, this choice corresponds to maximal $CP$-violation ($|\theta_{23}|=45^{\circ}, |\delta|=90^{\circ}$).

The mass matrix, Eq.~(\ref{retro}) with $y=z$, now, therefore, has an additional
symmetry: it commutes with the Mutativity ($\mu-\tau$ reflection) operator:
\beq
[\DMnsq,{\cal M}]=0
\label{commutator2}
\eeq
which is defined as multiplication by the $\mu-\tau$ exchange operator, followed by complex conjugation:
\begin{eqnarray}
{\cal M}={\cal C}P_{\mu\tau}, \quad{\rm where}\quad P_{\mu\tau}=\left( \matrix{  1  &  0  &   0  \cr
                                                                                 0  &  0  &   1  \cr
                                                                          0  &  1  &   0  \cr } \right)
\label{mtexch}
\end{eqnarray}
and ${\cal C}$ is the complex-conjugation operator.
We comment that the refined mass eigenstates (those of Eq.~(\ref{retro}), 
with $y=z$, which are given in the Appendix, Eq.~(\ref{Umns2})), as well 
as being eigenstates of Democracy, are now also eigenstates of Mutativity 
(as expected given Eq.~(\ref{commutator2})), with eigenvalues 
$|{\cal M}|=1$. Mutativity being an anti-unitary operator, the phases of 
the eigenvalues are not well-defined, and there is no corresponding 
conservation law. There are, however, phenomenological consequences of 
Mutativity symmetry (a symmetry which remains good, even in the presence 
of matter effects). These consequences have been explored thoroughly
in a previous publication \cite{mutativity} (independent, in fact, of the
constraint of Democracy).

The mixing matrix obtained assuming both Democracy and Mutativity is
displayed in Eq.~(\ref{Umns2}), and was first introduced in
Ref.~\cite{symmsGens}. It has a single unconstrained mixing angle,
$|U_{e3}|=\sin{\theta_{13}}\equiv\sqrt{\frac{2}{3}}\sin{\chi}$,
in addition to the two observable mass-squared differences,
$\DmsqS$ and $\DmsqA$. Expressions for these observables
in terms of the parameters are also given in the Appendix. However,
with the assumptions made so far, the $\nu_3$ mass eigenstate is not 
necessarily bi-maximally mixed (nor even nearly bi-maximally), as 
$|U_{e3}|$ remains unconstrained, and our mass matrix at this stage represents
only an intermediate step in our derivation. In fact, this mixing scheme 
interpolates between the earlier tri-bimaximal scheme \cite{tbm, hps4}, 
corresponding to the choice $\oddParam=0$ in Eq.~(\ref{retro}), yielding 
$\sin\theta_{13}=0$, and the original trimaximal scheme \cite{trimax}, 
having $x=y=z$, yielding $\sin\theta_{13}=1/\sqrt{3}$.

\vspace{5mm}
\ni {\bf 3 Chooz, Palo Verde and the Simplest Neutrino Mass Matrix}
\vspace{2mm}
\nl So far, the solar data, primarily, have motivated our introduction of the
Democracy symmetry, and the atmospheric data, that of the Mutativity symmetry. 
We still need a way to ensure a small value of $|U_{e3}|$, as required by
the Chooz and Palo Verde experiments. We continue in a similar vein, letting
the data be our guide. Assuming ${\cal D}$- and ${\cal M}$-invariance then, we consider the experimental constraints on the values of the three remaining parameters, $x$, $y$ and $\oddParam$. We find (after inverting Eqs.~(\ref{chi2}), (\ref{dmsqa2}) and (\ref{dmsqs2})):
\begin{eqnarray}
x&=&\frac{\DmsqA}{3}(2\sin^2\chi-3/2)+\frac{\DmsqS}{3}\simeq-\frac{\DmsqA}{2} \label{xx}\\
y\,(=z)&=&-\frac{\DmsqA}{3}\sin^2\chi+\frac{\DmsqS}{3}=\frac{\DmsqA}{3}(\frac{\DmsqS}{\DmsqA}-\sin^2\chi)\label{yy}\\
\oddParam&=&\frac{\DmsqA}{2}\sin 2\chi,
\label{oddPar}
\end{eqnarray}
where $\sin^2\chi=(3/2)\sin^2\theta_{13}<0.06$, as indicated by the
results of the Chooz and Palo Verde experiments \cite{reactor}, and
$(\DmsqS/\DmsqA)=0.027\pm 0.010$ \cite{dmsq}. Evidently, $y\,(=z)<<x$, while
$\oddParam<0.4\,x$.

It is clear that the parameter $\oddParam$ is at least somewhat smaller than $x$, and that the parameter $y\,(=z)$ is ``very small'' in the sense that it is considerably smaller than the upper-limit on $\oddParam$. There is no obvious reason why $y\,(=z)$ should be so small (it follows from the smallness of both $\sin^2\chi$ and $(\DmsqS/\DmsqA)$, as indicated by experiment).
We would therefore like to consider the phenomenology of the case in which the very small parameter $y\,(=z)$ is, in fact, exactly zero.
This gives six ``texture'' zeroes \cite{ross} in the real part of the (reduced) mass-squared matrix, and results in only two free parameters, $x$ and $\oddParam$, remaining:
\begin{eqnarray}
\DMnsq=x
\left( \matrix{  1  &  0  &   0  \cr
                 0  &  0  &   1  \cr
                 0  &  1  &   0  \cr } \right) + i\frac{\oddParam}{\sqrt{3}}
\left( \matrix{  0  &  1  &  -1  \cr
                -1  &  0  &   1  \cr
                 1  & -1  &   0  \cr } \right).
\label{simplest}
\end{eqnarray}
While the constraint $y\,(=z)=0$ does not obviously increase the symmetry of $\DMnsq$ as a whole, it certainly increases that of the real term alone, although it is not clear what significance, if any, this has. We will proceed anyway to consider the phenomenology, noting that our equations will be at least approximately valid, as long as $y(=z)<<d$.

By construction, the mass-squared matrix, Eq.~(\ref{simplest}), still has a single
trimaximally mixed mass-eigenstate (by Democracy), and $|U_{\mu 3}|=|U_{\tau 3}|$
(by Mutativity), but now also builds-in a relationship between the experimentally
small mixing parameter, $|U_{e3}|=\sin{\theta_{13}}\,(=\sqrt{\frac{2}{3}}\sin{\chi})$,
and the experimentally small mass-squared difference ratio, $\DmsqS/\DmsqA$.
In particular, setting $y\,(=z)=0$ in Eq.~(\ref{yy}), we see that\footnote{This prediction was first made, somewhat obliquely, in \cite{symmsGens}}:
\begin{equation}
\sin^2{\chi}=\DmsqS/\DmsqA
\label{ssqchi}
\end{equation}
which yields the new prediction:
\begin{equation}
\sin{\theta_{13}}=\sqrt{2\DmsqS/3\DmsqA}\simeq 0.13\pm 0.03,
\label{prediction}
\end{equation}
clearly consistent with the constraints from the Chooz and Palo Verde
experiments \cite{reactor} (see also eg.~\cite{deGouvea} for similar, but distinct relations based on different reasoning). The smallness of $\sin{\theta_{13}}$ is
now ensured and hence, the $\nu_3$ eigenstate is nearly bi-maximally
mixed, as required.

The observables and parameters are related as follows:
\begin{eqnarray}
\DmsqS&=&x\pm\sqrt{\oddParam^2+x^2}\label{dmsqs3}\\
\DmsqA&=&\pm2\sqrt{\oddParam^2+x^2}\label{dmsqa3}\\
\tan{2\chi}&=&\frac{\oddParam}{x}\label{oddPar3}.
\end{eqnarray}
For a normal neutrino mass hierarchy, the positive square-root is taken,
and $x$ is negative. For an inverted hierarchy, take the negative square-root and $x$ positive. The MNS lepton mixing matrix is given by
Eq.~(\ref{Umns2}), with:
\begin{equation}
\sin^2{\chi}=\frac{1}{2}[1-\frac{x}{(d^2+x^2)^{\frac{1}{2}}}], \quad
\cos^2{\chi}=\frac{1}{2}[1+\frac{x}{(d^2+x^2)^{\frac{1}{2}}}].
\label{sinchi}
\end{equation}
From Eq.~(\ref{ssqchi}), we also have $\cos^2{\chi}=\DmsqAmS/\DmsqA$, where
$\DmsqAmS=\DmsqA-\DmsqS$. We can now re-write Eq.~(\ref{simplest}) succinctly in terms of observables:
\begin{equation}
\DMnsq=\frac{\DmsqA}{2}[\cos{2\chi}\,P_{\mu\tau}+\frac{i}{\sqrt{3}}\sin{2\chi}\,\Aa]
\label{simplest2}
\end{equation}
where $P_{\mu\tau}$ and $\Aa$ are the first and second
matrices appearing in Eq.~(\ref{simplest}), respectively.

We believe this ansatz (or texture) to be the simplest possible
one for the neutrino mass-squared matrix, consistent with the data, and
admitting $CP$ violation. Certainly, the
neutrino mass matrix needs an overall scale, and at least one small
parameter, making a minimum of two parameters (there could be other
two-parameter schemes, but it is difficult to imagine any with fewer).
The six observables of neutrino
oscillations include 3 large (in fact, $\sim$maximal) mixing angles
($\theta_{12}$, $\theta_{23}$ and $\delta$), one smaller mixing angle
($\theta_{13}$ -- we know only its upper limit, at present), one
very small mass-squared difference ratio, $\DmsqS/\DmsqA$, and one
overall mass scale, $\DmsqA$, say. These are parameterised here, by a
mass matrix, Eq.~(\ref{simplest}), characterised simply by an overall
mass-scale, $\sqrt{\oddParam^2+x^2}=\DmsqA/2$, and a single small ratio,
$(\oddParam/\sqrt{3}x)\simeq 0.2$ (we note that $d\simeq\sqrt{\DmsqS\DmsqA}\simeq 0.34x$). The three large mixing angles are
thus fixed by the Democracy and Mutativity symmetries, while the
small and very small observables are respectively, $\sim$ linear and
$\sim$ quadratic in the small ratio, $(\oddParam/\sqrt{3}x)$:
\beq
\sin^2\theta_{13}=(2\DmsqS/3\DmsqA)\simeq\frac{1}{2}({\oddParam}/{\sqrt{3}x})^2.
\eeq
While we cannot say why the ratio $(\oddParam/\sqrt{3}x)$ is small,
we are left needing to explain the smallness of only this single quantity.

\vspace{5mm}
\ni {\bf 4 Predictions for Long Baseline Neutrino Experiments}
\vspace{2mm}
\nl Our expression for $\sin{\theta_{13}}$, Eq.~(\ref{prediction}), enables us to predict appearance and survival probabilities in future long-baseline neutrino experiments. The leading term in the electron-neutrino appearance probability at eg.~a neutrino superbeam experiment would be \cite{huber}:
\begin{eqnarray}
P(\nu_{\mu}\to\nu_e)&\simeq& 2 \sin^2{\theta_{13}}\cos^2{\theta_{13}}\sin^2\left(\frac{\DmsqA L}{4E}\right)\\
&=&(0.035\pm0.013)\sin^2\left(\frac{\DmsqA L}{4E}\right)
\end{eqnarray}
($L$ is the propagation-distance of the neutrino), while the leading term in the reactor anti-neutrino disappearance probability would be:
\begin{eqnarray}
1-P(\nu_{\overline{e}}\to\nu_{\overline{e}})&\simeq& 4 \sin^2{\theta_{13}}\cos^2{\theta_{13}}\sin^2\left(\frac{\DmsqA L}{4E}\right)\\
&=&(0.070\pm0.026)\sin^2\left(\frac{\DmsqA L}{4E}\right)
\end{eqnarray}
(it is exactly double $P(\nu_{\mu}\to\nu_e)$ above, as a consequence of
Mutativity symmetry). Thus, observable consequeneces are significant,
and, in particular, hold out the possibility of the measurement of
$\theta_{13}$ in the next generation of long-baseline neutrino
oscillation experiments \cite{huber}.

Looking further ahead to the prospects for the observation of $CP$ violation in neutrino oscillations, we note that because the smallness of the solar mass-squared  difference is linked to the smallness of $|U_{e3}|$ in this scenario, in the limit of $CP$ conservation ($\sin{\theta_{13}}\rightarrow 0$), $\DmsqS\rightarrow 0$
also. Hence, we see that the well-known suppression of $CP$ violation in neutrino oscillations by two factors, namely the smallness of $\DmsqS$ and that of $|U_{e3}|$ \cite{CPatNF}, is attributable to the single small $CP$-violating parameter $(\oddParam/\sqrt{3}x)$, on which the leading $CP$-violating term has a $\sim$ cubic dependence. This suppression is, of course, somewhat compensated in our present scenario, by the largeness of the other three mixing parameters, $\theta_{12}$, $\theta_{23}$ and $\delta$, leaving open the possibility to observe $CP$ violation in neutrino oscillations at some point in the future.

\vspace{5mm}
\ni {\bf 5 Discussion}
\vspace{2mm}
\nl In this simplest of mixing schemes then, all the
features of neutrino oscillation experiments (aside from the LSND
result \cite{lsnd}) are manifest.
Both terms of Eq.~(\ref{simplest}) are separately symmetric under
both the Democracy and Mutativity operators.
Taking the two terms in order of
decreasing magnitude, one may consider the leading term, proportional to
$x$, to {\it explain} the atmospheric data, its magnitude being
directly related to the atmospheric scale, Eq.~(\ref{xx}), and, on
its own, being diagonalised by a simple $2\times 2$ maximal mixing matrix 
operating in the $\mu-\tau/2-3$ subspace, thereby ensuring 
$|\sin\theta_{23}|=1/\sqrt{2}$. Such a mass matrix taken in isolation,
however, would result in a degeneracy between the $\nu_1$
and $\nu_2$ mass eigenstates, and zero values for the remaining mixing 
parameters.

The second term in Eq.~(\ref{simplest}), proportional to $\oddParam$, can be thought of as a perturbation, lifting the $\nu_1-\nu_2$ degeneracy with a small eigenvalue difference, $\DmsqS \simeq \oddParam^2/2x$, and collapsing the $\nu_2$ eigenstate into the trimaximally mixed state, Eq.~(\ref{nu2}). It
therefore {\it explains} the solar data, $\DmsqS$, and
$|U_{e2}|=1/\sqrt{3}$. The interplay between the two terms in Eq.~(\ref{simplest})
ensures a small $\sin\theta_{13}$, of order $\oddParam/x$, allowing $CP$ to 
be violated.

We comment finally, that we have not considered here the nature of the neutrino
as a Majorana or a Dirac particle. Our considerations apply equally well in
either case. By considering the reduced mass-squared matrix, $\DMnsq$ of Eq.~(\ref{retro}),
in which we have removed the dependence on the parameter $a$ of Eq.~(\ref{s3group}),
we have ensured no dependence on the overall offset of the neutrino masses.
The addition to $\DMnsq$ in Eq.~(\ref{retro}) of any multiple of the identity, to represent such an offset (thereby recovering the full $M^2_{\nu}$), clearly respects all the symmetries considered here. Our analysis is therefore both insensitive to and unpredictive with respect to the phenomenology of neutrinoless double beta decay.

\newpage
\vspace{7mm}
\ni \boldmath{\bf Acknowledgements}\unboldmath
\vspace{2mm}
\nl This work was supported by the UK Particle Physics and Astronomy
Research Council (PPARC). One of us (PFH) acknowledges the hospitality
of the Centre for Fundamental Physics (CFFP) at CCLRC Rutherford Appleton
Laboratory.

\vspace{5mm}
\newpage
\ni {\bf Appendix: Relationships Between Parameters and Observables}
\vspace{2mm}
\nl In order to keep the exposition of the main concepts of this paper 
concise, we provide some of the more complex formulae in this Appendix.

We start with the most general Democracy-invariant mass-squared matrix, Eq.~(\ref{s3group}).
In terms of the parameters $a$, $b$, $x$, $y$ and $z$, the eigenvalues may be written \cite{characters}:
\begin{eqnarray}
m_1^2 & = & a-{\rm Re} \, b 
              - \sqrt{ \, 3 \, ({\rm Im} \, b)^2 
                                +[(x-y)^2+(y-z)^2+(z-x)^2]/2} \label{m1sq} \\
m_2^2 & = & a + 2 {\rm Re} \, b +x+y+z \label{m2sq} \\ 
m_3^2 & = & a-{\rm Re} \, b 
              + \sqrt{ \, 3 \, ({\rm Im} \, b)^2 
                                +[(x-y)^2+(y-z)^2+(z-x)^2]/2} \label{m3sq}
\end{eqnarray}
and the MNS lepton mixing matrix is \cite{characters}:
\begin{eqnarray}
     \matrix{  \hspace{0.4cm} \nu_1 \hspace{1.8cm}
               & \hspace{0.5cm} \nu_2 \hspace{1.7cm}
               & \hspace{0.5cm} \nu_3  \hspace{0.5cm} }
                                      \hspace{2.0cm} \nonumber \\
\hspace{0.3cm}U \hspace{0.3cm} = \hspace{0.3cm}
\matrix{ e \hspace{0.2cm} \cr
         \mu \hspace{0.2cm} \cr
         \tau \hspace{0.2cm} }
\left( \matrix{    \sqrt{\frac{2}{3}} c_{\chi} c_{\phi} 
                 +i\sqrt{\frac{2}{3}} s_{\chi}s_{\phi}  &
                      \frac{1}{\sqrt{3}} &
                 -\sqrt{\frac{2}{3}} c_{\chi}s_{\phi}
   -i\sqrt{\frac{2}{3}} s_{\chi} c_{\phi} \cr 
   -  \frac{c_{\chi}c_{\phi}+is_{\chi}s_{\phi}}{\sqrt{6}}  
     -\frac{c_{\chi}s_{\phi}-is_{\chi}c_{\phi}}{\sqrt{2}} &
          \frac{1}{\sqrt{3}} &
    -\frac{c_{\chi}c_{\phi}-is_{\chi}s_{\phi}}{\sqrt{2}}  
     +\frac{c_{\chi}s_{\phi}+is_{\chi}c_{\phi}}{\sqrt{6}}  \cr
      \hspace{2mm}
     - \frac{c_{\chi}c_{\phi}+is_{\chi}s_{\phi}}{\sqrt{6}}  
     +\frac{c_{\chi}s_{\phi}-is_{\chi}c_{\phi}}{\sqrt{2}}
                                                   \hspace{2mm} &
         \hspace{2mm}
         \frac{1}{\sqrt{3}} \hspace{2mm} &
     \,\,\,\,\,\,\,\frac{c_{\chi}c_{\phi}-is_{\chi}s_{\phi}}{\sqrt{2}}  
     +\frac{c_{\chi}s_{\phi}+is_{\chi}c_{\phi}}{\sqrt{6}} 
                                       \hspace{2mm} \cr } \right)
\label{Umns}
\end{eqnarray}
where we have used the abbreviations:
$c_{\chi}= \cos \chi$, $s_{\chi}= \sin \chi$,
$c_{\phi}= \cos \phi$, $s_{\phi}= \sin \phi$,
and the phenomenological parameters $\chi$ and $\phi$ are given by:
\begin{eqnarray}
\tan 2 \phi & = & \frac{\sqrt{3} \, (z-y)}{(z-x)+(y-x)}, \label{phi}\\
\tan 2 \chi & = & \frac{\sqrt{6} \: {\rm Im \, b}}
               {[ \, (x-y)^2+(y-z)^2+(z-x)^2 \, ]^\frac{1}{2}} \label{chi}.
\end{eqnarray}
The usual mixing observables in the standard (PDG \cite{PDG}) parameterisation 
may be written in terms of $\chi$ and $\phi$:
\begin{eqnarray}
\sin^2{\theta_{13}}&=&\frac{1}{3}(1-\cos{2\chi}\cos{2\phi}) \label{s13sq} \\
\sin^2{\theta_{12}}&=&\frac{1}{3\cos^2{\theta_{13}}} \label{s12sq} \\
\sin^2{\theta_{23}}&=&\frac{1}{2}-\frac{\cos{2\chi}\sin{2\phi}}{2\sqrt{3}\cos^2{\theta_{13}}}
\label{s23sq}
\end{eqnarray}
and the $CP$-violation parameter $J$ \cite{jarlskog} is given by:
\begin{equation}
J = \frac{\sin 2\chi}{6\sqrt{3}} 
  = \frac{{\rm Im} \: b}{3\sqrt{2} \, [ \, 6 \, ({\rm Im} \, b)^2 
                       +(x-y)^2+(y-z)^2+(z-x)^2 \, ]^\frac{1}{2}} 
\label{jcp},
\end{equation}
from which the $CP$-violating phase $\delta$ can be extracted if required.
It is manifest that all observables are invariant if we add a given real 
constant to $a$ and $b$, providing that we subtract the same constant from 
$x$, $y$ and $z$. Of course all mass-squared differences are independent of $a$.

Without loss of generality then, we can choose ${\rm Re} \, b=0$, and hence 
we can write the two mass-squared differences:
\begin{eqnarray}
\DmsqA:=m_3^2-m_1^2 & = & 2 \sqrt{ \, \oddParam^2 
                                +[(x-y)^2+(y-z)^2+(z-x)^2]/2} \label{dmsqa} \\
\DmsqS:=m_2^2-m_1^2 & = & x+y+z \nonumber \\
                    & + & \sqrt{ \, \oddParam^2 
                                +[(x-y)^2+(y-z)^2+(z-x)^2]/2} \label{dmsqs}
\end{eqnarray}
where $\oddParam=\sqrt{3}~{\rm Im} \: b$. 
The expressions for the mixing parameters, Eqs.~(\ref{Umns})-(\ref{jcp}), remain unaltered.

Invoking Mutativity, ie.~setting $y=z$, Eq.~(\ref{phi}) immediately gives $\tan{2\phi}=0$
(and hence $\sin{2\phi}=0$), so that (we choose $\sin{\phi}=0$) the mixing matrix is:
\begin{eqnarray}
     \matrix{  \hspace{0.4cm} \nu_1 \hspace{0.6cm}
               & \hspace{0.5cm} \nu_2 \hspace{0.9cm}
               & \hspace{0.5cm} \nu_3  \hspace{0.5cm} }
                                      \hspace{0.5cm} \nonumber \\
\hspace{0.3cm}U \hspace{0.3cm} = \hspace{0.3cm}
\matrix{ e \hspace{0.2cm} \cr
         \mu \hspace{0.2cm} \cr
         \tau \hspace{0.2cm} }
\left( \matrix{    \sqrt{\frac{2}{3}}c_{\chi}  &
                      \frac{1}{\sqrt{3}} &
   -i\sqrt{\frac{2}{3}} s_{\chi} \cr 
      -\frac{c_{\chi}}{\sqrt{6}}  
      +\frac{is_{\chi}}{\sqrt{2}} &
          \frac{1}{\sqrt{3}} &
     -\frac{c_{\chi}}{\sqrt{2}}  
     +\frac{is_{\chi}}{\sqrt{6}}  \cr
      \hspace{2mm}
     -\frac{c_{\chi}}{\sqrt{6}}  
     -\frac{is_{\chi}}{\sqrt{2}}
                                                   \hspace{2mm} &
         \hspace{2mm}
         \frac{1}{\sqrt{3}} \hspace{2mm} &
     \,\,\,\,\,\,\,\frac{c_{\chi}}{\sqrt{2}}  
     +\frac{is_{\chi}}{\sqrt{6}} 
                                       \hspace{2mm} \cr } \right)
\label{Umns2}
\end{eqnarray}
where
\begin{eqnarray}
\tan 2 \chi & = & \frac{\oddParam}
               {(x-y)}.
\label{chi2}
\end{eqnarray}
Substituting in Eqs.~(\ref{s13sq})-(\ref{jcp}) we find that this corresponds to standard parameters:
\begin{eqnarray}
\sin^2{\theta_{13}}&=&\frac{2}{3}\sin^2{\chi} \label{s13} \\
\sin^2{\theta_{12}}&=&\frac{1}{3\cos^2{\theta_{13}}} \label{s12} \\
\sin^2{\theta_{23}}&=&\frac{1}{2} \label{s23} \\
\sin^2{\delta}&=&1. \label{del}
\end{eqnarray}
We also have:
\begin{eqnarray}
\DmsqA:=m_3^2-m_1^2 & = & 2 \sqrt{ \, \oddParam^2
                                +(x-y)^2} \label{dmsqa2} \\
\DmsqS:=m_2^2-m_1^2 & = & x+2y \nonumber \\
                    & + & \sqrt{ \, \oddParam^2
                                +(x-y)^2} \label{dmsqs2}.
\end{eqnarray}

Finally, setting $y\,(=z)=0$, we arrive at the simplest mass matrix, as discussed in the main text, and in particular, Eqs.~(\ref{dmsqs3})-(\ref{oddPar3}) above.

\end{document}